\documentclass[journal,transmag]{IEEEtran}
\usepackage{xcolor}
\usepackage{graphicx}
\usepackage{cite}
\usepackage{amsmath}
\usepackage{algorithmic}
\usepackage{array}
\usepackage[caption=false,font=footnotesize]{subfig}
\usepackage{fixltx2e}
\usepackage{dblfloatfix}
\usepackage{url}

\begin{document}
%
\title{Periodic phase diagrams in micromagnetics with an eigenvalue solver}


\author{\IEEEauthorblockN{Fangzhou Ai\IEEEauthorrefmark{1, 2},
Zhuonan Lin\IEEEauthorrefmark{2, 3},
Jiawei Duan\IEEEauthorrefmark{2, 3}, and
Vitaliy Lomakin\IEEEauthorrefmark{1, 2, 3},~\IEEEmembership{Fellow,~IEEE}}
\IEEEauthorblockA{\IEEEauthorrefmark{1}Department of Electrical and Computer Engineering, University of California, San Diego, La Jolla, CA 92093, U.S.A}
\IEEEauthorblockA{\IEEEauthorrefmark{2}Center for Memory and Recording Research, University of California, San Diego, La Jolla, CA 92093, U.S.A}
\IEEEauthorblockA{\IEEEauthorrefmark{3}Program in Materials Science and Engineering, University of California, San Diego, La Jolla 92093, U.S.A}
\thanks{Manuscript received ****; revised ****. 
Corresponding author: V. Lomakin (email: vlomakin@ucsd.edu).}}

\markboth{IEEE Transactions on Magnetics, manuscript}%
{Ai \MakeLowercase{\textit{et al.}}: IEEE Transactions on Magnetics Journals}
%



\IEEEtitleabstractindextext{%
\begin{abstract}
This work introduces an approach to compute periodic phase diagram of micromagnetic systems by solving a periodic linearized Landau-Lifshitz-Gilbert (LLG) equation using an eigenvalue solver with the Finite Element Method formalism. The linear operator in the eigenvalue problem is defined as a function of the periodic phase shift wave vector. The dispersion diagrams are obtained by solving the eigenvalue problem for complex eigen frequencies and corresponding eigen states for a range of prescribed wave vectors. The presented approach incorporates a calculation of the periodic effective field, including the exchange and magnetostatic field components. The approach is general in that it allows handling 3D problems with any 1D, 2D, and 3D periodicities. The ability to calculated periodic diagrams provides insights into the spin wave propagation and localized resonances in complex micromagnetic structures.
\end{abstract}

\begin{IEEEkeywords}
Periodic phase diagram, micromagnetics, eigenvalue solver
\end{IEEEkeywords}}

\maketitle

\IEEEdisplaynontitleabstractindextext

%
\IEEEpeerreviewmaketitle

\section{Introduction}
%
%
%
%

\IEEEPARstart{T}{he} study of micromagnetic systems is important for our understanding of the fundamental behaviors of magnetic materials and their applications. At the core of these studies lies the Landau-Lifshitz-Gilbert (LLG) equation, a fundamental tool for describing the dynamic evolution of magnetization in response to internal and external excitations. However, analyzing the dynamics of periodic micromagnetic structures remains a significant computational challenge due to the complexity of boundary conditions and the interactions between periodic elements.

Periodic systems, such as magnetically large object and patterned nanostructures, exhibit unique properties arising from their symmetry and periodic boundary conditions. The linearized form of the LLG equation \cite{10.1063/9.0000609} provides a pathway to analyze the stability, resonance phenomena, and spin wave propagation in such systems, but it necessitates an accurate and efficient calculation of the periodic fields. Conventional periodic micromagnetic methods often encounter difficulties in handling the periodicity or with phase shift\cite{WANG201084, fdtd, WYSOCKI2017274, PhysRevB.87.174422}, leading to challenges in the ability to calculate phase diagrams and eigenvalue spectra in general 3D structures with 1D, 2D, and 3D periodicities\cite{PhysRevLett.85.2817, PhysRevB.69.174401, PhysRevB.83.054434}.

\begin{figure}[!t]
\centering
\includegraphics[width=2.5in]{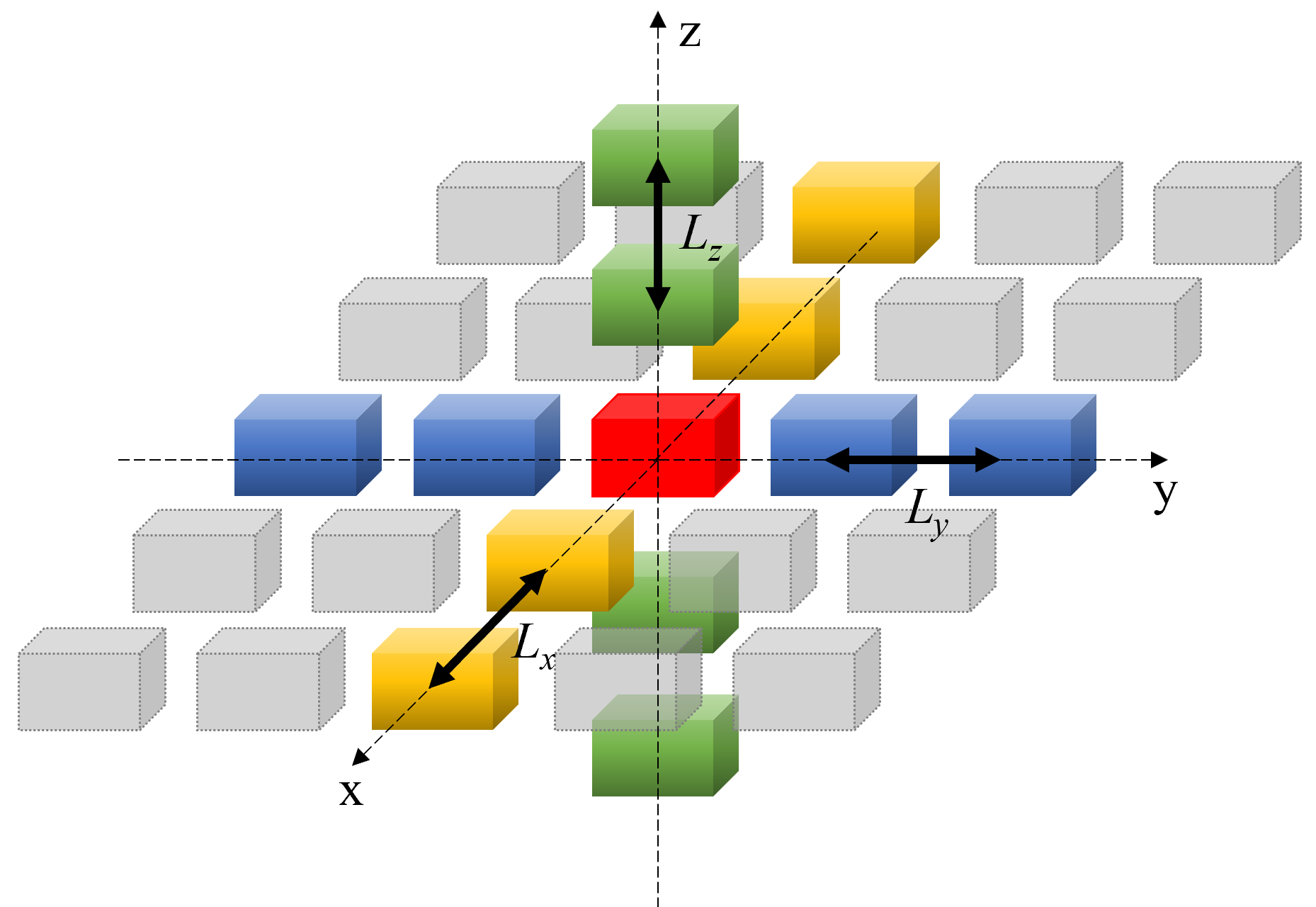}
\caption{Illustration for a periodic problem consisting along $\hat{x}$ (yellow), $\hat{y}$ (blue) and $\hat{z}$ (green) direction with zero-th unit cell (red) and its periodic images (grey).}
\label{fig:geo}
\end{figure}

In this paper, we propose an approach to solve eigenvalue problems using the linearized LLG equation with periodic boundary conditions in the Finite Element Method (FEM) framework. We consider general 3D periodic problems possibly with 1D, 2D, and 3D periodicities (Fig.~\ref{fig:geo}). A key aspect of our methodology is the precise computation of the periodic fields, which ensures compatibility with the micromagnetic framework while preserving the periodicity of the system. By employing this approach, we aim to construct periodic phase diagrams that capture the stability and dynamic (spin wave propagation) properties of micromagnetic systems under various conditions.

We validate our method through applications to well-studied periodic micromagnetic systems, demonstrating its efficiency and accuracy in calculating the periodic problem with phase shift. We then demonstrate the power of the approach by calculating dispersion diagrams of 1D and 2D periodic problems. This work contributes a new computational framework for analyzing periodic micromagnetic structures, offering potential insights for researchers in dynamic magnetism, spintronics, and materials science.

\section{Problem formulation}

We consider a 3D domain (Fig. 1) with the a magnetic structure of an arbitrary shape that is infinitely periodic in 1D, 2D, or 3D directions with periodicities of $L_x$, $L_y$, and $L_z$, respectively. We develop a formulation that allows calculating the dispersion diagrams for the periodically modulated magnetization with the same periodicites as the domain. 

This section presets the formulation of the periodic linearized LLG equation and field with phase-shifted periodic boundary condition (PBC), leading to the ability to set up a periodic eigenvalue problem.

\subsection{Periodic linearized LLG equation}

The magnetization space and time dependence is governed by the non-linear LLG equation:
\begin{equation} \label{eq:llg}
\frac{\partial \mathbf{M}}{\partial t} = -\frac{\gamma}{1+\alpha^2}\left(\mathbf{M}\times \mathbf{H}_\text{eff} + \frac{\alpha}{M_s}\mathbf{M}\times\mathbf{M}\times\mathbf{H}_\text{eff}\right),
\end{equation}
where $\gamma$ is the gyromagnetic ratio, $\alpha$ is the damping constant, $M_s$ is the saturation magnetization, and 
\begin{equation} \label{eq:heff}
\mathbf{H}_\text{eff} = \mathbf{H}_\text{ex} + \mathbf{H}_\text{ms} + \mathbf{H}_\text{an} + \mathbf{H}_\text{ap}
\end{equation}
is the effective magnetic field that includes the exchange, magnetostatic, anisotropy, and applied field components. The effective field also can include additional components, such as spin transfer torque \cite{RALPH20081190}, spin orbit torque \cite{shao2021roadmap}, eddy currents \cite{hrkac2005three}, magnetostriction \cite{ANDRA20018892}, etc.

Under the assumption of a weak excitation, the magnetization can be considered as a weak perturbation $\mathbf{v}$ around the  equilibrium state $\mathbf{M}_0$, i.e., $\mathbf{M}=\mathbf{M}_0+\mathbf{v}$. The effective field also can be written similarly: $\mathbf{H_{\text{eff}}}=\mathbf{H}_{\text{eff},0}+\mathbf{h}$, where $\mathbf{H}_{\text{eff},0}=\mathbf{H}_{\text{eff}}(\mathbf{M}_0)$ is the effective field for the magnetization equilibrium state and $\mathbf{h}=\mathbf{H}_{\text{eff}}(\mathbf{v})$ is the corresponding field perturbation. 

The equilibrium state satisfies Brown's condition $\mathbf{M}_0\times \mathbf{H}_0 = 0$ subject to a PBC with no periodic phase shift:
\begin{subequations} \label{eq:pv0}
    \begin{align} 
    \mathbf{M}_0&(\mathbf{r}+L_x\hat{\bf x}) =\mathbf{M}_0(\mathbf{r}),\\ 
    \mathbf{M}_0&(\mathbf{r}+L_x\hat{\bf x}+L_y\hat{\bf y}) =\mathbf{M}_0(\mathbf{r}),\\ 
    \mathbf{M}_0&(\mathbf{r}+L_x\hat{\bf x}+L_y\hat{\bf y}+L_z\hat{\bf z})=\mathbf{M}_0(\mathbf{r}).
\end{align}
\end{subequations}
for the 1D, 2D, or 3D periodicity, respectively. The equilibrium effective field $\mathbf{H}_{\text{eff},0}$ satisfies the same PBC with no phase shift as $\mathbf{M}_0$. The equilibrium state $\mathbf{M}_0$ is found by satisfying the Brown condition by energy minimization or solving the dynamic LLG equation\eqref{eq:llg} with high damping until no significant time variations are obtained.

The magnetization perturbation $\mathbf{v}$ satisfies the following phase-shifted PBC:
\begin{subequations} \label{eq:pv}
    \begin{align} 
    \mathbf{v}&(\mathbf{r}+L_x\hat{\bf x})=e^{-jk_{x0}L_x}\mathbf{v}(\mathbf{r}),\\ 
    \mathbf{v}&(\mathbf{r}+L_x\hat{\bf x}+L_y\hat{\bf y})=e^{-j\left(k_{x0}L_x+k_{y0}L_y\right)}\mathbf{v}(\mathbf{r}),\\ 
    \mathbf{v}&(\mathbf{r}+L_x\hat{\bf x}+L_y\hat{\bf y}+L_z\hat{\bf z})=e^{-j\left(k_{x0}L_x+k_{y0}L_y+k_{z0}L_z\right)}\mathbf{v}(\mathbf{r}).
\end{align}
\end{subequations}
for the 1D, 2D, or 3D periodicity, respectively. Here, $k_{x0}$, $k_{y0}$, and $k_{z0}$ are periodic phase shift wave numbers in the $x$, $y$, and $z$ direction, respectively. These phase shift wave numbers can be combined into a phase shift wave vector $\mathbf{k}_0=\hat{\bf x}k_{x0}+\hat{\bf y}k_{y0}+\hat{\bf z}k_{z0}$, which is defined depending on the dimensionality of the periodicity.

The perturbation field $\mathbf{h}$ satisfies the same PBC as $\mathbf{v}$, and it is linear in $\mathbf{v}$, i.e., it can be written as
\begin{equation} \label{eq:hv}
\mathbf{h} = \mathcal{C}(\mathbf{k}_0)\mathbf{v}
\end{equation}
where $\mathcal{C}(\mathbf{k}_0)$ is a linear operator that includes the linear operators corresponding to the effective field components. This operator depends on the wave vector $\mathbf{k}_0$.

Keeping only terms that are linear in the small perturbation $\mathbf{v}$ in the LLG equation Eq.~\eqref{eq:llg}, we obtain a linearized LLG equation \cite{PhysRevApplied.17.034016}
\begin{equation} \label{eq:lllg}
    \begin{aligned}
        \frac{d\mathbf{v}}{dt}=-&\frac{\gamma}{1+\alpha^2}\left(\mathbf{M}_0\times \mathbf{h}-\mathbf{H}_{\text{eff},0}\times\mathbf{v}\right) \\
        -&\frac{\gamma\alpha}{(1+\alpha^2)M_s}\mathbf{M}_0\times\left(\mathbf{M}_0\times \mathbf{h}-\mathbf{H}_{\text{eff},0}\times\mathbf{v}\right), \\
        =&{\mathcal A(\mathbf{k}_0)}\mathbf{v}
    \end{aligned}
\end{equation}
where ${\mathcal A}$ is a linear operator representing the right hand side of the equation and this operator accounts for the fact that the field perturbation $\mathbf{h}$ is linear with respect to $\mathbf{v}$ via \eqref{eq:hv}. The operator ${\mathcal A}$ is a function of the phases shift wave vector $\mathbf{k}_0$.

For finding the dispersion diagrams, we define $\mathbf{v}$ in the form of $\mathbf{v}(\mathbf{r},t)=e^{j\omega t}\varphi(\mathbf{r})$ and assume no external excitation, i.e., $\mathbf{H}_\text{ap}=0$, which allows writing the linearized equation \eqref{eq:lllg} as an eigenvalue problem:
\begin{equation} \label{eq:lllg_eigen}
    j\omega \varphi={\mathcal A}(\mathbf{k}_0)\varphi.
\end{equation}
Here, $\omega$ is the eigen frequency corresponding to the eigenstate $\varphi(\mathbf{r})$ that satisfies the PBC as in Eq.\eqref{eq:pv}. Since the operator ${\mathcal A}$ is a function $\mathbf{k}_0$, solving this eigenvalue problem result in calculating a dispersion diagram, i.e., the dependence of the eigen frequency of the wave vector $\mathbf{k}_0$. This dispersion diagram can be considered from two points of view. One can obtain a dependence of generally complex $\omega$ versus $\mathbf{k}_0$ by solving the explicit eigenvalue problem of Eq.~\eqref{eq:lllg_eigen} for a range of given real $\mathbf{k}_0$. Alternatively, one can obtain generally complex $\mathbf{k}_0$ for a set of given real $\omega$. The latter approach may require solving an implicit eigenvalue problem because $\mathbf{k}_0$ appears in the operator $\mathcal{A}$ implicitly. While being more complex problem to solve, this approach allows directly calculating not only the real but also imaginary parts of the wave numbers, thus providing the spin wave propagation length.

\subsection{Field under periodic boundary condition with phase shift}

Only the magnetostatic and exchange fields need special care in terms of PBCs since they come from non-localized interactions. For the exchange field, changes are needed when touching periodic boundary condition (T-PBC) is present, namely the object size is equal to the periodic length. 

The exchange fields corresponding to the dynamic perturbation is given by
\begin{equation} \label{eq:hex}
    \mathbf{h}_{\text{ex}}(\mathbf{r}) = \frac{2A_{ex}}{M_s^2(\mathbf{r})}\nabla^2 \mathbf{v}(\mathbf{r}).
\end{equation}
where $\nabla^2$ is the Laplacian operator and $A_{ex}$ is the exchange constant. In typical FEM in micromagnetics\cite{Fastmag}, the structure is meshed into a mesh, that often is based on tetrahedral tesselation. The solution is obtained as the magnetization states at the vertices of the mesh. Inside the mesh elements, the magnetization is interpolated via polynomials, which often are chose as linear. The Laplacian operator is implemented as a sparse matrix with the matrix band determined by the connectivity of the mesh vertices to the surrounding vertices connected via common elements \cite{Fastmag}. When the computational domain is smaller than the periodicity, there is no need to modify the conventional sparse matrix representation. When the computational domain extends through the periodic boundaries, the PBC of \eqref{eq:pv} needs to be accounted for by properly updating the sparse matrix. The exchange field corresponding to the equilibrium state $\mathbf{H}_{\text{ex,0}}$ is given by the right hand side of Eq.~\eqref{eq:hex}, where $\mathbf{v}$ is replaces with $\mathbf{M}_0$ and the PBC of Eqs.~\eqref{eq:pv0} is used.



The perturbation magnetostatic field $\mathbf{h}_{\text{ms}}$ is due to long-range interactions, and it can be calculated either by solving the Poisson equation or by evaluating the superposition integrals \cite{NUFFT, 10259661}. We evaluate the magnetostatic field using superposition integrals via the following formulation
\begin{subequations} \label{eq:hms}
    \begin{align}
        & \rho(\mathbf{r}) = \nabla\cdot\mathbf{M}(\mathbf{r}),~\rho_s(\mathbf{r})=-\hat{\mathbf{n}}\cdot\mathbf{M}(\mathbf{r}),\label{eq:hms1}\\
        & u(\mathbf{r})=\iiint_{V}G^p(\mathbf{\mathbf{r}-\mathbf{r'}})\rho(\mathbf{r})d\mathbf{r'}+\iint_{S}G^p(\mathbf{\mathbf{r}-\mathbf{r'}})\rho_s   (\mathbf{r})d\mathbf{r'},\label{eq:hms2}\\
        & \mathbf{H}_{\text{ms}}(\mathbf{r}) = -\nabla u(\mathbf{r}).\label{eq:hms3}
    \end{align}
\end{subequations}
Here, $\rho$ is the volumetric magnetic charge density, $\rho_s$ is the surface magnetic charge density, $u$ is the magnetic scalar potential, and $G_p$ is the 3D periodic Green's function (PGF), which can be 1D, 2D, or 3D periodic \cite{AI2024109291}. The gradient operator $\nabla$ is represented via a sparse matrix similar to the Laplacian operator as in Eq.~\eqref{eq:hex}. The term $G_p$ is the periodic Green's function (PGF) defined as 
\begin{subequations}
    \label{eq:2}
    \begin{align}
        G^{p}_{1D}(\mathbf{r}) = &\sum_{i_x=-\infty}^{\infty} e^{-jk_{x0}(i_xL_x)}G_0(\mathbf{r}-i_xL_x\mathbf{\hat{x}}),\\
        G^{p}_{2D}(\mathbf{r}) = &\sum_{i_x=-\infty}^{\infty}\sum_{i_y=-\infty}^{\infty}e^{-j[k_{x0}(i_xL_x)+k_{y0}(i_yL_y)]} \nonumber \\
        &G_0(\mathbf{r}-i_xL_x\mathbf{\hat{x}}-i_yL_y\mathbf{\hat{y}}),\\
        G^{p}_{3D}(\mathbf{r}) = &\sum_{i_x=-\infty}^{\infty}\sum_{i_y=-\infty}^{\infty}\sum_{i_z=-\infty}^{\infty}\nonumber \\
        &e^{-j[k_{x0}(i_xL_x)+k_{y0}(i_yL_y)+k_{z0}(i_zL_z)]}\nonumber \\
        &G_0(\mathbf{r}-i_xL_x\mathbf{\hat{x}}-i_yL_y\mathbf{\hat{y}}-i_zL_z\mathbf{\hat{z}}),
    \end{align}
\end{subequations}
for the 1D, 2D, and 3D periodicities, respectively, and $G_0$ is the free space scalar Green's function given by $G_0(\mathbf{r})=1/|\mathbf{r}|$. The integrals in Eq.~\eqref{eq:hms} can be calculated as a product of sparse matrixes representing the short range integrals related to the basis functions and short term superposition sums related to the long-range interactions \cite{Fastmag}. The long range superposition sum can be efficiently handled by the Fast Fourier Transform periodic interpolation method\cite{AI2024109291, FPIP, Derek} with the periodic Green's function (PGF).

\section{Results}

This section demonstrates results obtained using the periodic eigenvalue LLG solver. The structure is meshed via a tetrahedral mesh and linear nodal elements are used. The dispersion diagrams are calculated by solving the eigenvalue problem for the eigen frequencies $\omega$ using the periodic eigenvalue LLG solver for a set of prescribed wave numbers. The results include a validation example and examples of calculating the dispersion diagrams for 3D problems with 1D and 2D periodicities.

We first validate the presented approach by calculating the dispersion relation of magnetostatic backward volume wave (MSBVW) in an infinitely large permalloy (Py) film.

\subsection{MSBVW dispersion relationship}

The theoretical dispersion relationship of the MSBVW is expressed as \cite{Stancil2009tq}
\begin{equation}
    \begin{aligned}
        \omega_n^2=&\lambda_{ex}\gamma M_s k^2 \times \left(\lambda_{ex}\gamma M_s k^2+\gamma M_s \frac{1-e^{-kd}}{kd}\right),
    \end{aligned}
\end{equation}
Here, the $\omega_n$ is the angular frequency of $n^{th}$ order spin wave, $k$ is the corresponding wave number,  $d$ is the thickness of the film, and $\lambda_{ex}$ is the exchange length defined as $\lambda_{ex}=2A_{ex}/M_s^2$.

\begin{figure*}[!t]
\centering
\subfloat{\includegraphics[width=2.8 in]{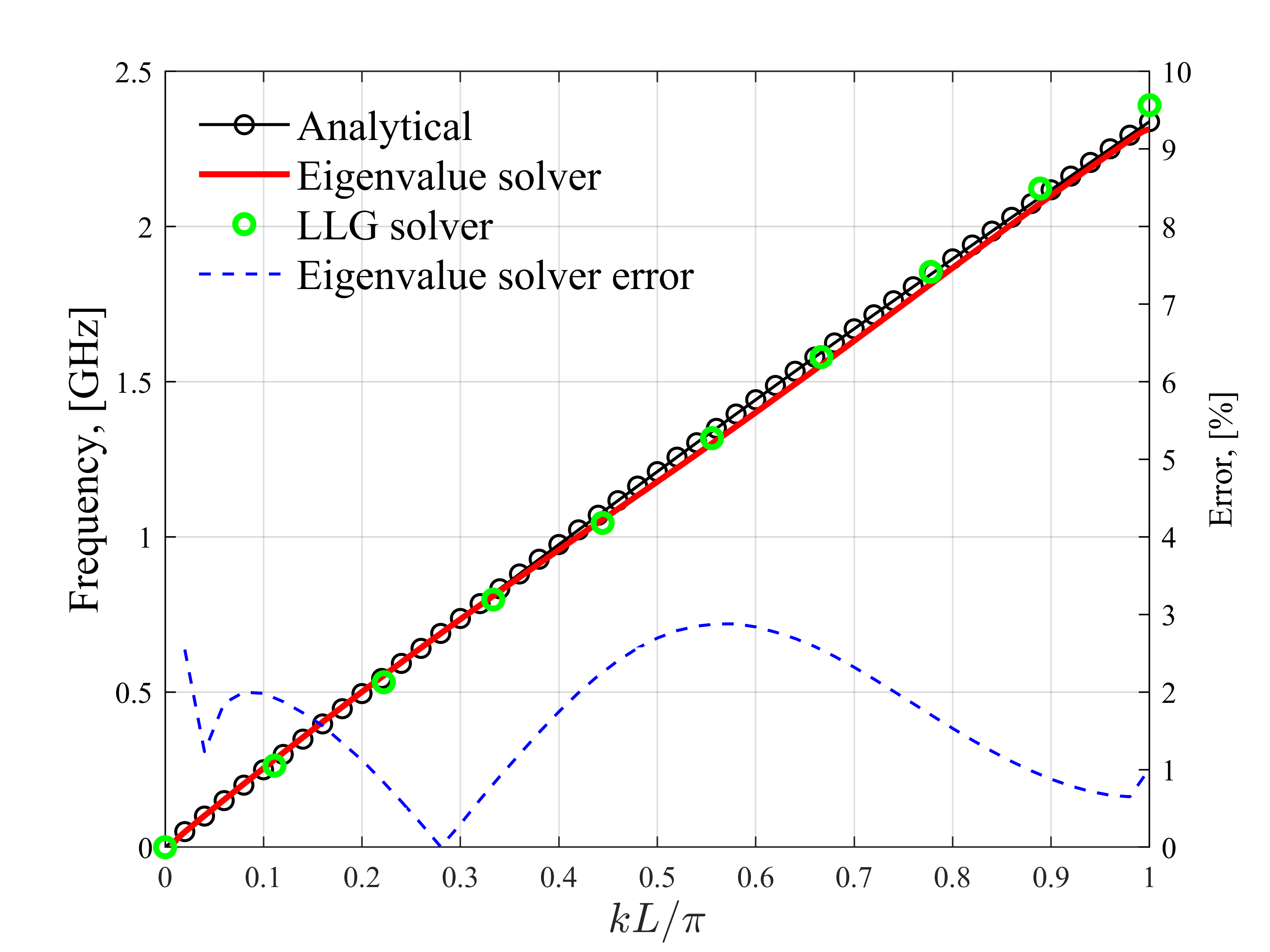}%
\label{fig_first_case}}
\hfil
\subfloat{\includegraphics[width=2 in]{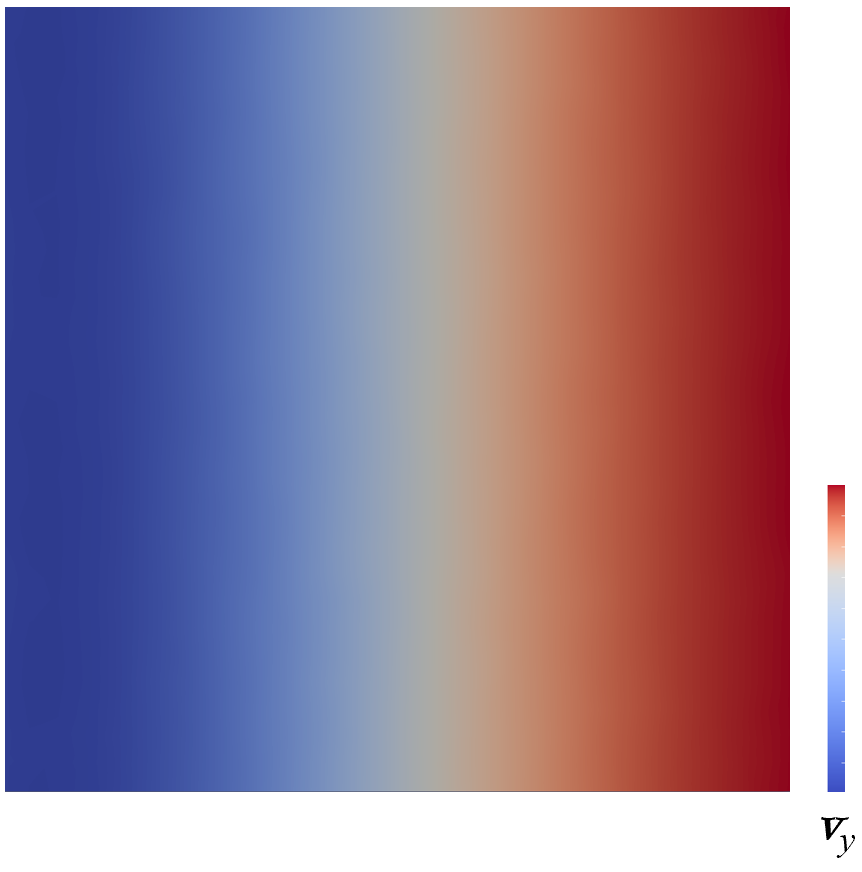}%
\label{fig_second_case}}
\caption{The left figure is the phase diagram (dispersion relationship) calculated from theoretical equation (black circles), LLG time-domain solver (green circles), eigenvalue solver (red line) and its relative error with respect to the theoretical values. The right figure is the $y$ component of the magnetization perturbation $\mathbf{v}$ from the eigenstate of $k_x=\pi/L_x$.}
\label{fig:msbvw}
\end{figure*}

In the numerical solution, we set the unit cell of size 200 $\times$ 200 $\times$ 20 nm, with the mesh edge length of 8 nm. The material parameters are $M_s=637~\text{emu/cc}$, $A_{ex}=1.4\times 10^{-6}~\text{erg/cm}$, no anisotropy, and $\alpha=0.02$. We impose a 2D PBC in the $x$ and $y$ directions, i.e., $L_x=L_y=200$ nm. The equilibrium magnetization is aligned along the $x-$direction. By sweeping $k=k_{x0}$ and keeping $k_{y0}=0$, the corresponding eigen-frequency related to the MSBVW of wave vector $k_x$ is calculated. For further comparison, we also calculate the dispersion relationship using the time-domain LLG solver. All these results are present in Fig.~\ref{fig:msbvw}, and the eigen state of the MSBVW with the wave vector $k=\pi/L_x$ is also plotted. We observe all methods agree well with each other validating the presented solver. We note that running the time domain LLG equation is much slower and has various issues, such as a possible non-linearity in the behavior.

\subsection{1D periodicity with a hole}

\begin{figure*}[!t]
\centering
\subfloat{\includegraphics[width=2.8 in]{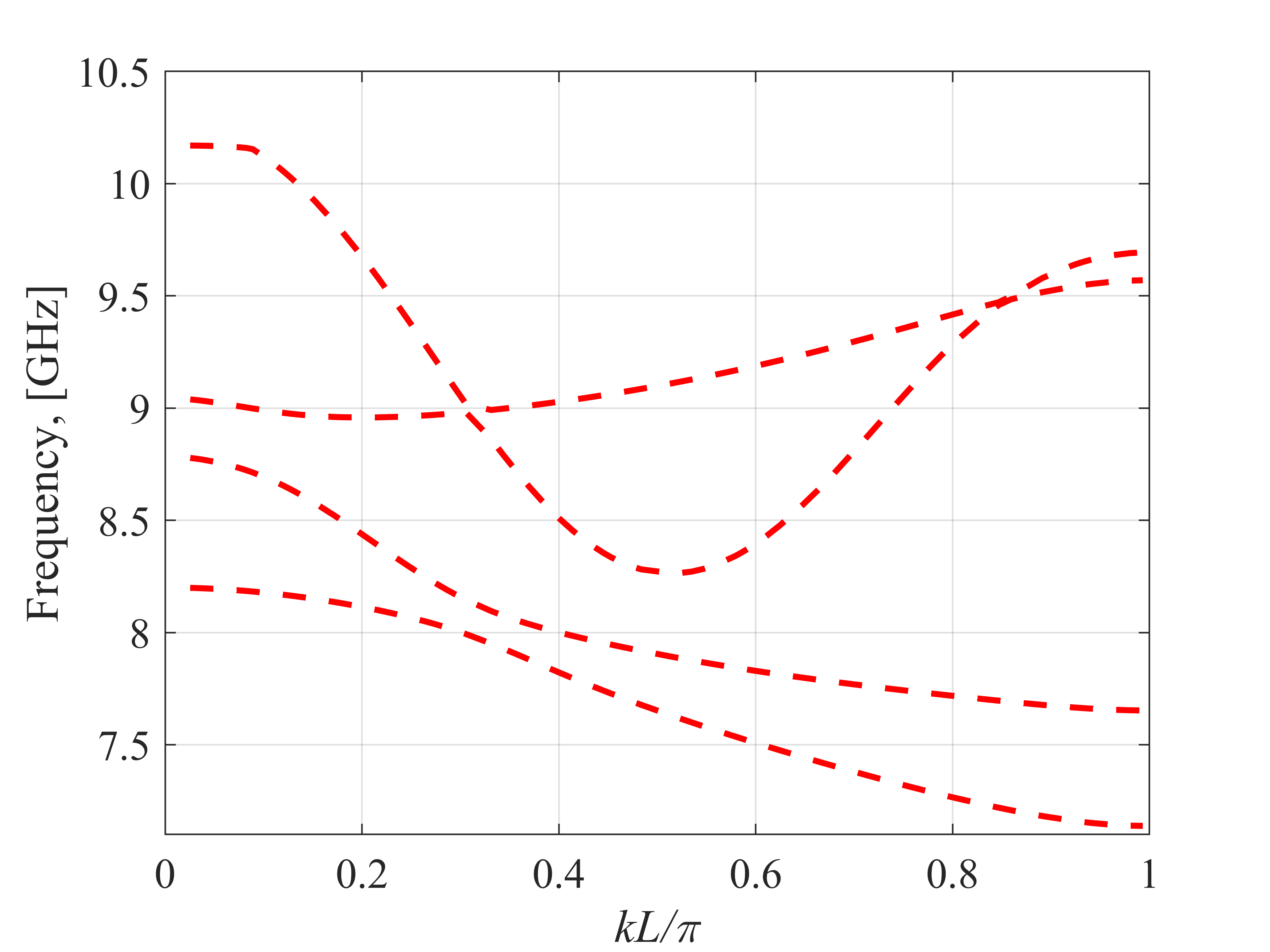}%
\label{fig_first_case}}
\hfil
\subfloat{\includegraphics[width=2 in]{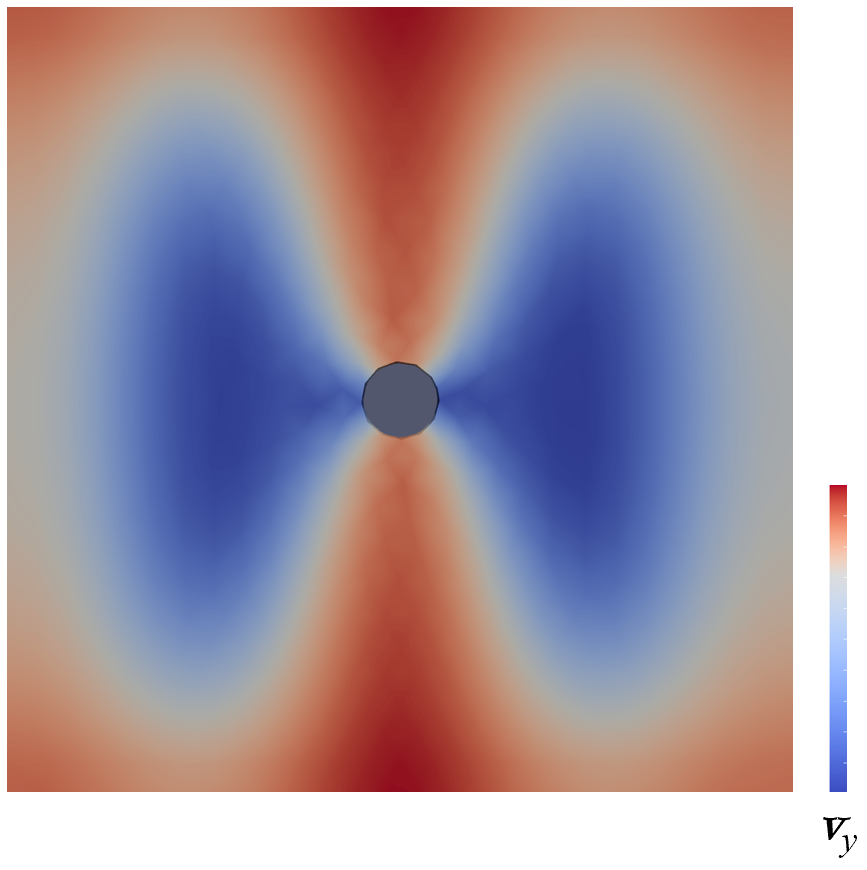}%
\label{fig_second_case}}
\caption{The left figure is the 1D phase diagram of the film with a hole in the middle from the periodic LLG eigen value solver (red dashed line). The right figure is the $y$-component of the magnetization perturbation $\mathbf{v}$ from the eigen state of $k_{x0}=\pi/L_x$.}
\label{fig:1d}
\end{figure*}

We then show a 1D periodic phase diagram for the same film with a periodic array of holes in the $x$ direction with the same periodicity of $200$nm, which is of a width of $200$nm in the $y$-direction. The equilibrium state, which is calculated via the periodic LLG time-domain solver\cite{ai2024periodicmicromagneticfiniteelement}, is found to be slightly different than that for the case of the uniform film. The eigen state of the lowest branch with $k_{x0}=\pi/L_x$ and the periodic phase diagram of the first 4 branches with the lowest energy are shown in Fig.~\ref{fig:1d}. The eigen state exhibits a butterfly feature which is very different from that in Fig.~\ref{fig:msbvw}.

\subsection{2D periodicity with a hole}
\begin{figure*}[!t]
\centering
\subfloat{\includegraphics[width=2.8 in]{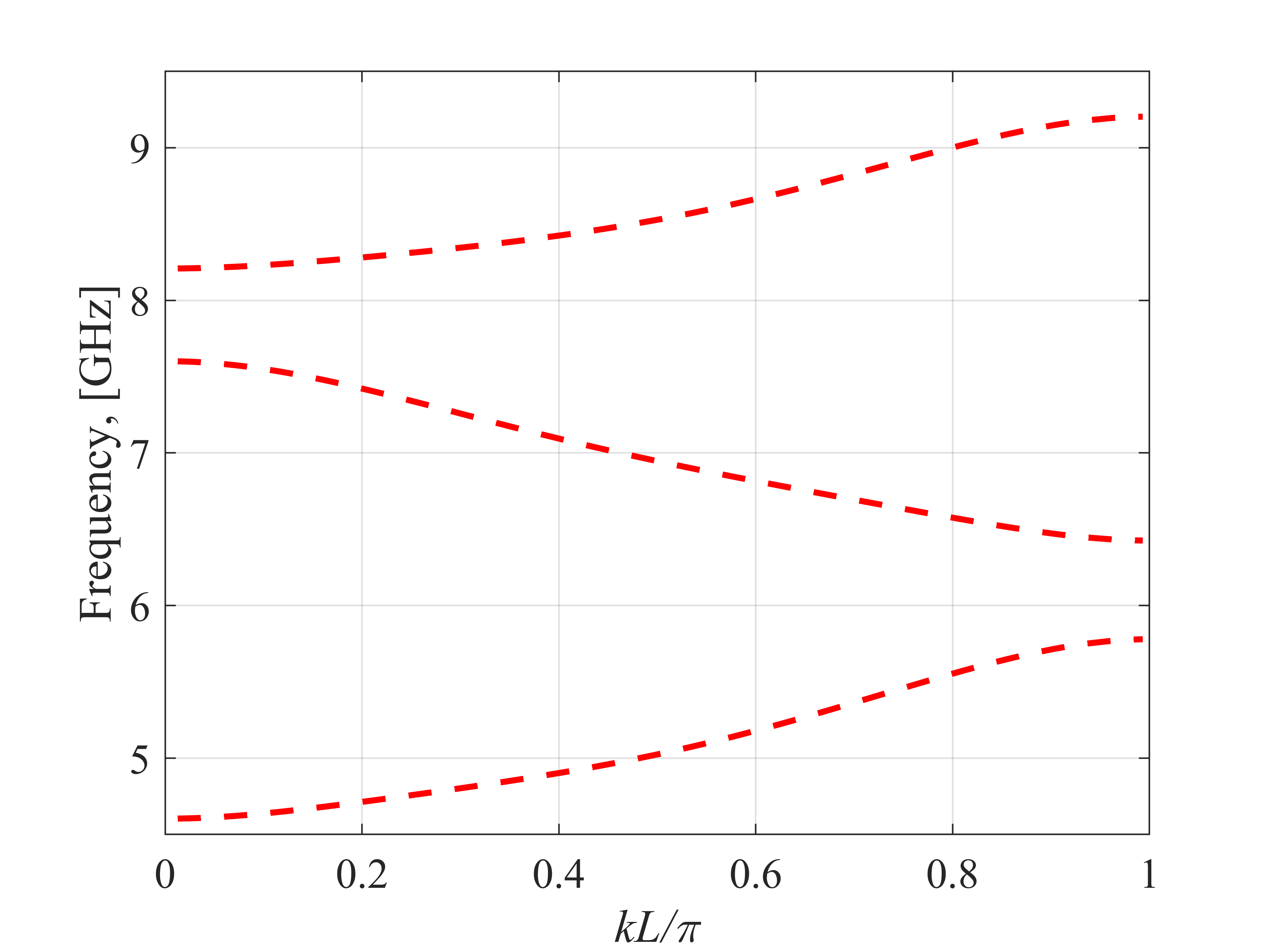}%
\label{fig_first_case}}
\hfil
\subfloat{\includegraphics[width=2 in]{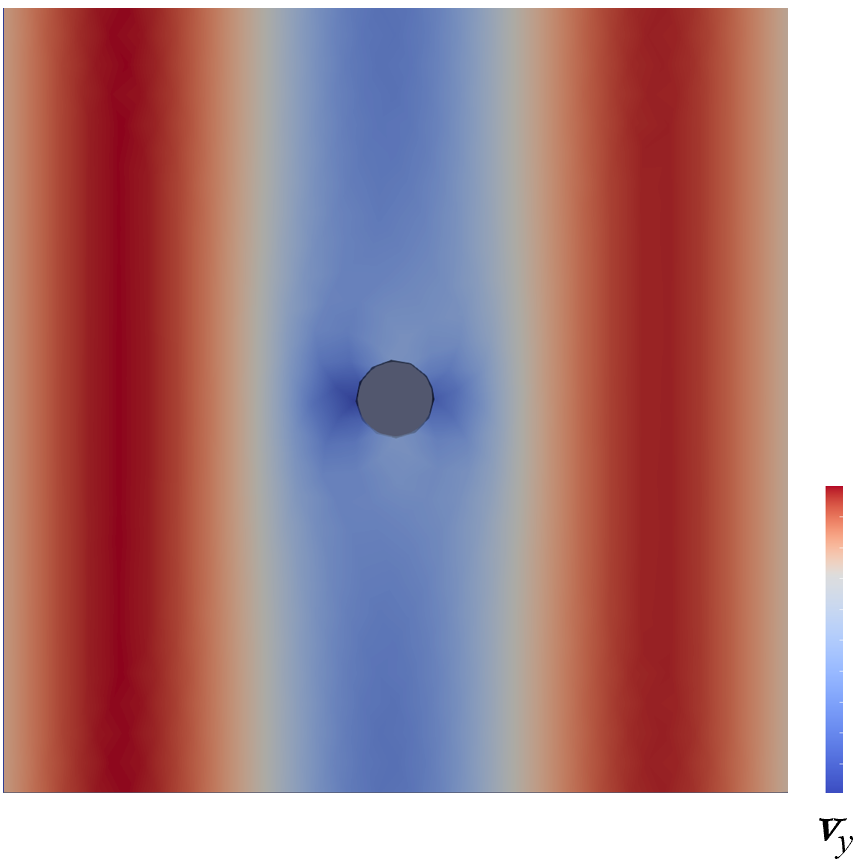}%
\label{fig_second_case}}
\caption{The left figure is the 2D phase diagram of the film with a hole in the middle from the periodic LLG eigenvalue solver (red dashed line). The right figure is the $y$ component of the magnetization perturbation $\mathbf{v}$ from the eigen state of $k_{x0}=\pi/L_x$.}
\label{fig:2d}
\end{figure*}

Finally, we calculate a periodic phase diagram of the above example with both $x$ and $y$ periodicities. The eigen state of the lowest branch with $k_{x0}=\pi/L_x$ and the periodic phase diagram of the first 3 branches with the lowest energy are shown in Fig.~\ref{fig:2d}. The eigenstate here is very different from that in Fig.~\ref{fig:msbvw}.

\section{Conclusion}
We presented a computational framework for calculating periodic phase diagrams in micromagnetic systems by solving a linearized LLG equation with an eigenvalue solver. A critical aspect of the presented approach is the calculation of the periodic field, which ensures the compatibility of PBCs with the micromagnetic formalism. By addressing the challenges associated with periodic systems, the presented approach offers an efficient and accurate means to analyze the stability and dynamic properties of micromagnetic structures.

The results demonstrate the validity and generality of the approach in capturing the key features of periodic micromagnetic systems, including resonance behaviors and phase behavior under varying conditions. Case studies illustrate the utility of the proposed method for investigating nanostructured materials and patterned magnetic systems, providing valuable insights into their dynamic properties.

This work not only advances the computational tools available for micromagnetic analysis but also lays the foundation for further exploration of periodic systems in nanomagnetism and materials science. Future studies could extend this methodology to include nonlinear dynamics, thermal effects, and more complex geometries, broadening its applicability to a wider range of micromagnetic phenomena.

\section*{Acknowledgment}
This work was supported in part by the Quantum Materials for Energy Efficient Neuromorphic-Computing (Q-MEEN-C), an Energy Frontier Research Center funded by the U.S. Department of Energy, Office of Science, Basic Energy Sciences under Award No. DESC0019273. The work was also supported in part by Binational Science Foundation, grant \#2022346. The work used Purdue Anvil cluster at Rosen Center for Advanced Computing (RCAC) in Purdue University through allocation ASC200042 from the Advanced Cyberinfrastructure Coordination Ecosystem: Services \& Support (ACCESS) program \cite{10.1145/3569951.3597559}, which is supported by National Science Foundation grants \#2138259, \#2138286, \#2138307, \#2137603, and \#2138296.
\bibliographystyle{IEEEtran}  
\bibliography{ref}   

\begin{IEEEbiographynophoto}{Fangzhou Ai}
received the B.S. degree in physics from Zhejiang University, Hangzhou, China in 2018, the M.s. and the Ph.D. degree in electrical and computer engineering from University of California, San Diego (UCSD), La Jolla, USA in 2020 and 2024. His research interests include high performance algorithms, GPU computing, and the study of magnetization dynamics phenomena.
\end{IEEEbiographynophoto}

\begin{IEEEbiographynophoto}{Zhuonan Lin}
received the B.S. degree from University of Science and Technology of China in 2015, the M.S. and Ph.D. degrees from University of California, San Diego in Materials Science and Engineering in 2017 and 2022.
\end{IEEEbiographynophoto}

\begin{IEEEbiographynophoto}{Jiawei Duan}
Received M.S. degree from University of California, San Diego in  Department of Electrical and Computer Engineering in 2021. Currently a Ph.D. candidate in Materials Science and Engineering under the supervision of Prof. Vitaliy Lomakin. His research interests include high performance computing and micromagnetic simulations. 
\end{IEEEbiographynophoto}

\begin{IEEEbiographynophoto}{Vitaliy Lomakin}
{Vitaliy Lomakin received his M.S. in Electrical Engineering from Kharkiv National University (Ukraine) in 1996 and Ph.D. in Electrical Engineering from Tel Aviv University (Israel) in 2003. From 2002 to 2005, he was a Postdoctoral Associate and Visiting Assistant Professor in the Department of Electrical and Computer Engineering, University of Illinois at Urbana Champaign. In 2005, he joined the University of California, San Diego, where he is Professor of Electrical and Computer Engineering. His research interests include the development of efficient computational techniques and analytical models for studying multiphysics problems, including Electromagnetics, Micromagnetics, and first principle simulations as well as using these techniques for the study of magnetic, optical, and microwave materials and devices.}
\end{IEEEbiographynophoto}
\end{document}